\documentclass[aps,prb,reprint,twocolumn,superscriptaddress,floatfix,citeautoscript]{revtex4-2}
\usepackage[bookmarks=true,colorlinks,linkcolor=OrangeRed,urlcolor=NavyBlue,citecolor=RoyalBlue]{hyperref}
\usepackage{epsfig,amsmath,amssymb,color,comment,physics}
\usepackage[makeroom]{cancel}
\usepackage[caption=false,font=normalsize]{subfig}
\usepackage{mathrsfs}
\usepackage[countmax]{subfloat}
\usepackage[normalem]{ulem}
\usepackage[english]{babel}
\usepackage{dsfont}
\usepackage{placeins}
\usepackage[dvipsnames]{xcolor}

\usepackage{multirow} 
\usepackage{graphicx}  

\usepackage{bbold} 
\usepackage{dcolumn}
\usepackage{bm}
\usepackage[capitalise]{cleveref}

\usepackage{tikz}
\usepackage{calc} 
\usepackage{wasysym} 

\begin{document}

\title{Temperature flow in pseudo-Majorana functional renormalization for quantum spins}

\author{Benedikt Schneider}
\affiliation{Department of Physics and Arnold Sommerfeld Center for Theoretical
Physics, Ludwig-Maximilians-Universit\"at M\"unchen, Theresienstr.~37,
80333 Munich, Germany}
\affiliation{Munich Center for Quantum Science and Technology (MCQST), 80799 Munich, Germany}
\author{Johannes Reuther}
\affiliation{Dahlem Center for Complex Quantum Systems and Institut f\"ur Theoretische Physik, Freie Universit\"{a}t Berlin, Arnimallee 14, 14195 Berlin, Germany}
\affiliation{Helmholtz-Zentrum f\"{u}r Materialien und Energie, Hahn-Meitner-Platz 1, 14109 Berlin, Germany}
\affiliation{Department of Physics and Quantum Centers in Diamond and Emerging Materials (QuCenDiEM) group, Indian Institute of Technology Madras, Chennai 600036, India}
\author{Mat\'ias G. Gonzalez}
\affiliation{Dahlem Center for Complex Quantum Systems and Institut f\"ur Theoretische Physik, Freie Universit\"{a}t Berlin, Arnimallee 14, 14195 Berlin, Germany}
\affiliation{Helmholtz-Zentrum f\"{u}r Materialien und Energie, Hahn-Meitner-Platz 1, 14109 Berlin, Germany}
\author{Bj\"orn Sbierski}
\affiliation{Department of Physics and Arnold Sommerfeld Center for Theoretical
Physics, Ludwig-Maximilians-Universit\"at M\"unchen, Theresienstr.~37,
80333 Munich, Germany}
\affiliation{Munich Center for Quantum Science and Technology (MCQST), 80799 Munich, Germany}
\affiliation{Institut für Theoretische Physik, Universit\"at T\"ubingen, Auf der Morgenstelle 14, 72076 T\"ubingen, Germany}

\author{Nils Niggemann}
\affiliation{Dahlem Center for Complex Quantum Systems and Institut f\"ur Theoretische Physik, Freie Universit\"{a}t Berlin, Arnimallee 14, 14195 Berlin, Germany}
\affiliation{Helmholtz-Zentrum f\"{u}r Materialien und Energie, Hahn-Meitner-Platz 1, 14109 Berlin, Germany}
\affiliation{Department of Physics and Quantum Centers in Diamond and Emerging Materials (QuCenDiEM) group, Indian Institute of Technology Madras, Chennai 600036, India}

\begin{abstract}
We implement the temperature flow scheme first proposed by Honerkamp and Salmhofer in Phys.~Rev.~B 64, 184516 (2001) into the pseudo-Majorana functional renormalization group method for quantum spin systems. Since the renormalization group parameter in this approach is a physical quantity -- the temperature $T$ -- the numerical efficiency increases significantly compared to more conventional renormalization group parameters, especially when computing finite temperature phase diagrams. 
We first apply this method to determine the finite temperature phase diagram of the $J_1$-$J_2$ Heisenberg model on the simple cubic lattice where our findings support claims of a vanishingly small nonmagnetic phase around the high frustration point $J_2=0.25J_1$. Perhaps most importantly, we find the temperature flow scheme to be advantageous in detecting finite temperature phase transitions as, by construction, a phase transition is never encountered at an artificial, unphysical cutoff parameter. Finally, we apply the temperature flow scheme to the dipolar XXZ model on the square lattice where we find a rich phase diagram with a large non-magnetic regime down to the lowest accessible temperatures. Wherever a comparison with error-controlled (quantum) Monte Carlo methods is applicable, we find excellent quantitative agreement with less than $5\%$ deviation from the numerically exact results.
\end{abstract}

\date{\today}
\maketitle

\section{Introduction}
\label{sec:introduction}
Frustrated quantum spin systems are known for their rich phenomenology, allowing for peculiar effects such as order-by-disorder phase transitions \cite{bergman2007order,mulder2010spiral} or magnetically disordered low-temperature phases \cite{wen2019experimental,Shastry1981,schafer2023abundance,lozano2023competing,Adhikary2021,pohle2023ground}. Examples for the latter are valence bond solids or the highly sought-after quantum spin liquids in which spin excitations fractionalize into emergent quasi-particles with unusual quantum statistics \cite{savary2016quantum}.
Given these desirable properties, there is a high demand for numerical techniques to treat these systems. While a few notable models are amenable to exact solutions \cite{kitaev2006anyons} or quantum Monte Carlo \cite{sandvik1991quantum,sandvik1999stochastic,benton2012seeing}, the treatment of general frustrated spin systems is notorious for its difficulty, even if significant approximations are employed.

In the last decade, the pseudo-fermion functional renormalization group (PFFRG) \cite{reuther_j_2010,muller2023pseudo} has established itself as a useful tool for the numerical treatment of a variety of spin models at zero temperature due to its remarkable flexibility \cite{chern2023pseudofermion,gresista2023candidate,noculak2023pseudofermion,niggemann_quantum_2023-1,fukui2023ground,Keles2022}. Treating spin operators through a fermionic particle representation to leverage the established functional renormalization group formalism, the PFFRG does not suffer from the sign problem and can treat translational invariant systems with arbitrary lattice geometry and two-body spin interactions. One of its shortcomings, the inclusion of unphysical states in the pseudo-fermion representation, can be circumvented at finite temperatures via the Popov-Fedotov trick \cite{schneider_taming_2022,popov1988functional} or by avoiding unphysical states altogether via a faithful spin representation in terms of Majorana fermions in the pseudo-Majorana (PM)-FRG \cite{niggemann_frustrated_2021,niggemann_quantitative_2022}. It was found that these approaches accurately capture the interplay between thermal and quantum fluctuations, enabling computations of magnetic phase diagrams and critical temperatures in quantitative agreement to quantum Monte Carlo (whenever the latter is applicable). On the other hand, while PMFRG simulations at finite temperature represent an important method extension and become even perturbatively error-controlled at large temperature, the cost of this improvement is significant, as it requires a separate solution of the numerically expensive FRG flow equations at each temperature. The renormalization group parameter is typically implemented through an artificial infrared cutoff $\Lambda$ in the single particle Green function, suppressing fermionic propagation with Matsubara frequencies $|\omega_n| \ll \Lambda$. 
An alternative formulation was first demonstrated by Honerkamp and Salmhofer in Ref.~\cite{honerkamp_temperature-flow_2001} for systems of itinerant fermions where {\it temperature} was employed as a flow parameter instead. In this approach, a single FRG flow along the physical temperature provides a whole slice through a finite-temperature phase diagram at once. On the other hand, however, the usual notion of RG as a successive integration of UV degrees of freedom is lost \cite{Platt2013,Honerkamp2004}.

Motivated by the above advantages, in this paper, we demonstrate an implementation of the temperature flow scheme in the context of the PMFRG. Specifically, in \cref{sec:Rescaling} we explain the key methodological step in this approach which amounts to defining rescaled fields and vertex functions. We present flow equations for the rescaled vertices and observables in terms of the rescaled fields in \cref{sec:FlowEq}. Formally, this temperature flow formulation of the FRG corresponds to an independent method, whose results in the strong coupling limit can differ from the conventional scheme using a Matsubara frequency cutoff (hereafter referred to as $\Lambda$-flow scheme). Hence, we first benchmark our new method on the Heisenberg model on the simple cubic lattice in \cref{sec:J1J2} against error-controlled quantum Monte Carlo and explain crucial differences between the $\Lambda$-flow and temperature flow ($T$-flow) schemes in \cref{sec:discussion_transition}. Next, in \cref{sec:Dipolar} we demonstrate an investigation of a more application-oriented frustrated spin system of high current research interest. Specifically, in \cref{sec:Dipolar} we treat the square lattice dipolar XXZ model amenable to experimental realizations~\cite{christakisProbingSiteresolved2023,chenContinuousSymmetryBreaking2023} where we exploit the efficiency of the temperature flow formalism in determining the finite temperature phase diagram. Finally, we summarize our findings in \cref{sec:Conclusion}.

\section{Action and field rescaling}
\label{sec:Rescaling}
We assume a general spin-1/2 Hamiltonian 
\begin{align}\label{eq:ham_general}
    H = \sum_{i,\alpha}h^\alpha_i S^\alpha_i +\frac{1}{2}\sum_{i,j,\alpha_1,\alpha_2}S^{\alpha_1}_i J^{\alpha_1\alpha_2}_{ij}S^{\alpha_2}_j,
\end{align}
where $S_i^\alpha$ with $\alpha=x,y,z$ are the components of a spin-1/2 operator on site $i$, $J_{ij}^{\alpha\beta}$ are general anisotropic spin interactions and $h_i^\alpha$ is a site-dependent magnetic field. We map $H$ onto a pseudo-Majorana Hamiltonian using the $SO(3)$ representation \cite{martin1959generalized,tsvelik_new_1992} 

\begin{align}
 S^x_i = -i\eta^y_i\eta^z_i,\ S^y_i = -i\eta^z_i\eta^x_i,\ S^z_i = -i\eta^x_i\eta^y_i.   \label{eq:MajoranaRep}
\end{align}
This representation's main advantage is the fact that it does not feature unphysical states and thus we may proceed without the need of any projection.

To solve the corresponding Majorana Hamiltonian, we first consider a general system of interacting Majoranas with the action written in imaginary time $\tau$ \cite{niggemann_frustrated_2021}
\begin{align}
    S&=\frac{1}{2}\int_0^\beta \dd \tau \; \eta_{\alpha_1}(\tau)(\delta_{\alpha_1\alpha_2}\partial_\tau+iA_{\alpha_1 \alpha_2})\eta_{\alpha_2}(\tau) \nonumber\\
    &+ \frac{1}{4!} \int_0^\beta \dd  \tau \;V_{\alpha_1\alpha_2 \alpha_3 \alpha_4}\eta_{\alpha_1}(\tau)\eta_{\alpha_2}(\tau)\eta_{\alpha_3}(\tau)\eta_{\alpha_4}(\tau) .\label{eq:ActionS0}
\end{align}

Here, Einstein summation is assumed, $\beta = 1/T$ is the inverse temperature and $\eta_\alpha(\tau)$ are real and antisymmetric Majorana fields  satisfying $\{\eta_\alpha,\eta_\beta\} = \delta_{\alpha\beta}$  and $\eta_\alpha(\tau)^\dagger = \eta_\alpha(-\tau)$, while $\alpha$ refers to indices labeling an arbitrary set of single-particle quantum numbers.
 The key step to derive a temperature flow FRG scheme is to gather all temperature dependence in the non-interacting part of the Hamiltonian. Here, we do this by introducing a modified Fourier transform 
 \begin{align}
     \eta(\omega) &=T^{\frac{1}{4}}\int^1_0 \dd \tau \; e^{-i\omega\tau}\left(\frac{\tau}{T}\right),\nonumber\\
     \eta\left(\frac{\tau}{T}\right)&=T^{-\frac{1}{4}}\sum_\omega  e^{i\omega\tau}\eta(\omega),
 \end{align}
  with the dimensionless Matsubara frequencies  $\omega=\pi(2n+1)$ and $n \in \mathbb{Z}$, which is more convenient but otherwise equivalent to the rescaling of Majorana fields as done by Honerkamp and Salmhofer 
 \cite{honerkamp_temperature-flow_2001}. Crucially, the transformation is chosen in such a way that no implicit temperature dependencies enter through frequencies and the interacting part of the rescaled action.
This way, we may express Eq.~\eqref{eq:ActionS0} as
\begin{align}
S = -\frac{1}{2}\sum_{\substack{\omega_1,\omega_2 \\\alpha_1\alpha_2}}&\eta_{\alpha_1}(\omega_1) G^{-1,T}_{0;\alpha_1\alpha_2}(\omega_1,\omega_2)\eta_{\alpha_2}(\omega_2)
\label{eq:ActionRescaled}\\+\frac{1}{4!}\sum_{\substack{\omega_1,\ldots,\omega_4\\\alpha_1 \dots \alpha_4}}&V_{\alpha_1 \alpha_2 \alpha_3 \alpha_4}\delta_{\omega_1+\omega_2+\omega_3+\omega_4,0}\nonumber\\
\times &\eta_{\alpha_1}(i\omega_1)\eta_{\alpha_2}(i\omega_2)\eta_{\alpha_3}(i\omega_3)\eta_{\alpha_4}(i\omega_4)
\end{align}

where we defined
\begin{equation}
    G^{-1,T}_{0;\alpha_1\alpha_2}(\omega_1,\omega_2) =  \frac{i}{\theta(T)}[\omega_1\delta_{\alpha_1,\alpha_2}-\theta(T)^2A_{\alpha_1 \alpha_2}]\delta_{\omega_1,-\omega_2} \label{eq:G0}
\end{equation} as the bare Green function. The crucial insight is that $\theta(T)=T^{-\frac{1}{2}}$ can be seen as a regulator function since it implies a vanishing propagator $G^T_0 \rightarrow 0$ for $T\xrightarrow{}\infty$. In the usual FRG formalism this is achieved by a regulator $G_0 \rightarrow \Theta^\Lambda G_0$ where the function $\Theta^\Lambda$ vanishes at the start of the flow at $\Lambda\rightarrow\infty$. We note that, while this suppression does not by itself act as an infrared cutoff of the Matsubara frequencies, the finite temperature has a similar effect of regularizing infrared divergencies as it shifts the smallest Matsubara frequency away from zero.
In \cref{eq:ActionRescaled}, the temperature dependence is fully contained in the regulator $\theta(T)$, which trivially generates the same hierarchy of flow equations as in the standard FRG formalism (see, for example Ref.~\cite{kopietz_introduction_2010}), upon simply replacing all derivatives with respect to the artificial cutoff $\Lambda$ by derivatives with respect to $T$.

\section{Flow Equations and observables}
\label{sec:FlowEq}
\subsection{General flow equations}
The FRG flow equations are derived from the action [Eq. \eqref{eq:ActionRescaled}] in full analogy to the standard PMFRG formalism \cite{niggemann_frustrated_2021}.
In this fermionic language, 
 Majorana Green functions are defined as
the bare propagator $G^T_0$, full propagator $G^T$ and connected two particle Green function $G^{4,T}_c$
\begin{align}
G^T_{1,2} &=\langle\eta_2\eta_1\rangle, \\
G^{4,T}_{c;1,2,3,4} &=\langle\eta_4\eta_3\eta_2\eta_1\rangle-\langle\eta_4\eta_3\rangle\langle\eta_2\eta_1\rangle \nonumber \\ & +\langle\eta_4\eta_2\rangle\langle\eta_3\eta_1\rangle-\langle\eta_3\eta_2\rangle\langle\eta_4\eta_1\rangle, 
\end{align}
where we have introduced the superlabels $1 = (i_1,\mu_1,\omega_1)$ that collectively describe site, spin and frequency index where the latter emerges after Fourier transforming the associated imaginary-time ordered correlation functions. 

In the FRG formalism, the objects of interest are the self-energy $\Sigma_{1,2}$ and the four-point vertex $\Gamma_{1,2,3,4}$ which are related to Green functions via the Dyson equation and the tree expansion \cite{kopietz_introduction_2010}
\begin{align}
    \Sigma^T_{1,2} &= G^{-1,T}_{0;1,2}-G^{-1,T}_{1,2},\\
    \Gamma^T_{1,2,3,4} &= -\sum_{1^\prime,\ldots,4^\prime}
    G^{-1,T}_{1,1^\prime}
    G^{-1,T}_{2,2^\prime} 
    G^{-1,T}_{3,3^\prime} 
    G^{-1,T}_{4,4^\prime} 
    G^{4,T}_{c;1^\prime,2^\prime,3^\prime,4^\prime}.
\end{align}
As outlined in Ref.~\cite{niggemann_frustrated_2021}, in thermal equilibrium the Green functions and vertices are frequency conserving while due to a  local $\mathbb{Z}_2$ gauge symmetry in the Majorana representation [Eq.~\eqref{eq:MajoranaRep}] the propagator and self energy are local and the vertex is bi-local, 

\begin{align}
    G^T_{1,2} =& G^T_{i_1;\alpha_1\alpha_2}(\omega_2)\delta_{i_1,i_2}\delta_{\omega_1,-\omega_2},\\
    G^{-1,T}_{0;1,2} =& G^{-1,T}_{0;i_1;\alpha_1\alpha_2}(\omega_1)\delta_{i_1,i_2}\delta_{\omega_1,-\omega_2},\\
   \Sigma^T_{1,2}  =&\Sigma^{T}_{i_1;\alpha_1\alpha_2}(\omega_1)\delta_{i_1,i_2}\delta_{\omega_1,-\omega_2},\\
    \Gamma^T_{1,2,3,4} =& \delta_{\omega_1+\omega_2+\omega_3+\omega_4,0}[\\ &\Gamma^T_{i_1i_3;\alpha_1\alpha_2\alpha_3\alpha_4}(\omega_1,\omega_2,\omega_3,\omega_4)\delta_{i_1,i_2}\delta_{i_3,i_4}\nonumber \\
    -&\Gamma^T_{i_1i_2;\alpha_1\alpha_3\alpha_2\alpha_4}(\omega_1,\omega_3,\omega_2,\omega_4)\delta_{i_1,i_3}\delta_{i_2,i_4}\nonumber \\
    +&\Gamma^T_{i_1i_2;\alpha_1\alpha_4\alpha_2\alpha_3}(\omega_1,\omega_4,\omega_2,\omega_3)\delta_{i_1,i_4}\delta_{i_2,i_3}].\nonumber
\end{align}

In the following, we provide the flow equations for the interacting free-energy $f_\textrm{int} = F_\textrm{int}/N = -T\log (\frac{Z}{Z_0})$, where $N$ is the number of sites, 
the self-energy $\Sigma^{T}_{i_1;\alpha_1,\alpha_2}(\omega_1)$ and vertex $\Gamma^{T}_{i_1 i_3;\alpha_1 \alpha_2 \alpha_3 \alpha_4}(\omega_1,\omega_2,\omega_3,\omega_4)$ that can be derived equivalently to Ref.~\cite{niggemann_frustrated_2021}.
With the transfer frequencies 
\begin{align}
    s & = \omega_1 + \omega_2 = -\omega_3 - \omega_4,\\
    t & = \omega_1 + \omega_3 = -\omega_2 - \omega_4,\\
    u & = \omega_1 + \omega_4 = -\omega_2 - \omega_3
\end{align}
these flow equations are given by
\begin{widetext}
\begin{align}
    \frac{\dd}{\dd T}\frac{f_\textrm{int}}{T}
    &= \frac{1}{2N}\sum_{k}\sum_{\omega} \sum_{\beta_1 \dots \beta_4}\left(\frac{\partial}{\partial T} G^{T}_{0;k;\beta_{1}\beta_{2}}(\omega) \right)\Sigma^{T}_{k;\beta_{2}\beta_{3}}(\omega)G^{T}_{k;\beta_{3}\beta_{4}}(\omega)G^{-1,T}_{0;k;\beta_{4}\beta_{1}}(\omega),\label{eq:FlowEqG0}
    \\
    \frac{\dd}{\dd T}\Sigma^{T}_{i;\alpha_{1}\alpha_{2}}(\omega)
    &=
    \frac{1}{2} \sum_{k}\sum_{\omega^\prime} \sum_{\beta_1 \beta_2}\Gamma_{ki;\beta_{2}\beta_{1}\alpha_{1}\alpha_{2}}^T(-\omega^\prime,\omega^\prime,\omega,-\omega)\frac{\partial}{\partial T}[G^T_{k;\beta_{1}\beta_{2}}(\omega^\prime)],
    \\
    \frac{\dd}{\dd T}\Gamma^T_{ij;\alpha_{1}\alpha_{2}\alpha_{3}\alpha_{4}}(s,t,u)
    &=
    X_{ij;\alpha_1,\alpha_2;\alpha_3,\alpha_4}(s,t,u)
    -
    \tilde{X}_{ij;\alpha_1,\alpha_3;\alpha_2,\alpha_4}(t,s,u) 
    + 
    \tilde{X}_{ij;\alpha_1,\alpha_4;\alpha_2,\alpha_3}(u,s,t),\label{eq:FlowEqVertex}
\end{align}
\begin{align}
    X_{ij;\alpha_1,\alpha_2;\alpha_3,\alpha_4} &= \frac{1}{2}\sum_{k,\omega} \sum_{\beta_1 \dots \beta_4}\Gamma_{ik;\alpha_{1}\alpha_{2}\beta_{1}\beta_{2}}^T(\omega_1,\omega_2,\omega-s,-\omega)\Gamma_{kj;\beta_{3}\beta_{4}\alpha_{3}\alpha_{4}}^T(\omega,s-\omega,\omega_3,\omega_4) P_{kk;\beta_{2}\beta_{3};\beta_{4}\beta_{1}}^{T}(\omega,\omega-s), \label{eq:X}\\
    \tilde{X}_{ij;\alpha_1,\alpha_2;\alpha_3,\alpha_4} &= \sum_{\omega} \sum_{\beta_1 \dots \beta_4}\Gamma_{ij;\alpha_{1}\beta_{1}\alpha_{3}\beta_{3}}^{T}(\omega_1,-\omega,\omega_2,\omega-s)\Gamma_{ij;\beta_{2}\alpha_{2}\beta_{4}\alpha_{4}}^{T}(\omega,\omega_3,s-\omega,\omega_4)P_{ij;\beta_{1}\beta_{2};\beta_{4}\beta_{3}}^{T}(\omega,\omega-s), \label{eq:Xtilde}
 \end{align}

\end{widetext}
where we define the single-scale propagator as
\begin{align} \frac{\partial}{\partial T} G^T_{k;\alpha_{1}\alpha_{2}}(\omega) =&- \sum_{\beta_1\beta_2}G^T_{k;\alpha_{1}\beta_{1}}(\omega)G^T_{k;\beta_{2}\alpha_2}(\omega) \nonumber\\
&\times  \left(\frac{\partial}{\partial T}G^{-1,T}_{0;k;\beta_{1}\beta_{2}}(\omega)\right),
\end{align}
and the bubble propagator as
\begin{equation}
     P_{ij;\alpha_{1}\alpha_{2};\alpha_{3}\alpha_{4}}^{T}(\omega,\omega-s) = \frac{\partial}{\partial T}\left[G_{i;\alpha_{1}\alpha_{2}}^{T}(\omega)G_{j;\alpha_{3}\alpha_{4}}^{T}(\omega-s)\right]. \label{eq:BubbleProp}
 \end{equation}

The main differences between the flow equations presented here and those of Ref.~\cite{niggemann_frustrated_2021} are the definition of the propagator and the absence of factors $T$ associated with the frequency sums.
The initial conditions follow immediately from the fact that the bare propagator $G_0^T$ vanishes at $T= \infty$ so that the only nonzero vertex at the beginning of the flow is the bare spin interaction:
\begin{align}
     \lim_{T\xrightarrow{}\infty}\frac{f_\textrm{int}}{T}&=0,\\
    \Sigma^{T\rightarrow \infty}_{i;\alpha_{1}\alpha_{2}}(\omega) &= 0,\\
     \Gamma^{T \rightarrow \infty}_{ij,\alpha_1\alpha_2\alpha_3\alpha_4}(s,t,u) &= -\sum_{\beta_1\beta_2}\epsilon_{\alpha_1\alpha_2\beta_1}J^{\beta_1\beta_2}_{ij}\epsilon_{\beta_2\alpha_1\alpha_2}, 
\end{align}
where $\epsilon_{\alpha_1\alpha_2\alpha_4}$ is the fully antisymmetric tensor. 
 Note that by convention the magnetic field is implemented in the off-diagonal elements of the bare inverse Green function in Eq.~\eqref{eq:G0} instead of in the self-energy with $A_{\alpha_1\alpha_2}$ given by
\begin{align}
    A_{\alpha_1\alpha_2} = -\sum_{\beta}\epsilon_{\alpha_1\alpha_2\beta}h^{\beta}.
\end{align}
In the Katanin truncation scheme \cite{katanin2004fulfillment} that we use for all calculations below, the partial derivative in Eq.~\eqref{eq:BubbleProp} is changed to a total derivative, thus including a feedback of the self-energy derivative into the vertex flow equation. This approximation, originally motivated by its inclusion of contributions from the six-point vertex, is known to dampen ordering tendencies which are otherwise overestimated \cite{muller2023pseudo}.

We emphasize that the self-energy defined above is related to the $\Lambda$-flow self-energy as $\Sigma^{\Lambda=0}(\omega) = T^{1/2}\Sigma^T(\omega)$, while the vertex is unchanged $\Gamma^{\Lambda=0}_{ij}(s,t,u) = \Gamma^{T}_{ij}(s,t,u)$~\footnote{Note that full the vertex function $\Gamma^\Lambda(\omega_1,\omega_2,\omega_3,\omega_4) \equiv \Gamma^\Lambda(s,t,u) \beta \delta_{\omega_1+\omega_2+\omega_3+\omega_4,0}$ has a relative factor of $\beta = 1/T$.}.
In practice, one can further reduce the number of independent vertex components by considering the spin and lattice symmetries of the model of interest \cite{niggemann_frustrated_2021,sbierski2023magnetism}.
\subsection{Observables}
A feature of the temperature flow is that we have direct access to the differentiated vertices with respect to temperature. Therefore, we have direct access to the free energy $f$ and mean energy $U=\langle H \rangle$ while the heat capacity $C = \frac{\dd U}{\dd T}$ can be obtained by numerical differentiation. By using the known result for the partition function of free spins-1/2 in a magnetic field $\bm h_i$
\begin{align}
    \log (Z_0) = \sum_i\log \left(2 \cosh \left(\frac{|\mathbf{h}_i|}{T}\right)\right)\label{eq:PartitionFunction0}
\end{align}
we can write them as 

\begin{align}
     f &=   f_\textrm{int} - T\log(Z_0),\label{eq:FreeEnergy}\\
      \frac{U}{N} &= -T^2 \frac{\dd}{\dd T}\left(\frac{f_\textrm{int}}{T} - \log(Z_0)\right),  \label{eq:MeanEnergy} \\
     C  &= \frac{\dd U}{\dd T}. \label{eq:HeatCapMain}
\end{align}

Other observables are the magnetization $M^\alpha_i=\langle S^\alpha_i\rangle$, magnetic susceptibility $\chi^{\alpha_1\alpha_2}_{ij}(\omega)=\int^\beta_0e^{i\omega\tau}\langle S_i^{\alpha_2}(\tau)S_j^{\alpha_1}(0)\rangle$ and the equal time spin-spin correlator $\langle S_i^{\alpha_2}S_j^{\alpha_1}\rangle$:
\begin{align}
    M^\alpha_j &=  -iT^{\frac{1}{2}}\sum_\omega\sum_{\beta_1\beta_2} \frac{\epsilon_{\alpha\beta_1\beta_2}}{2}G^T_{j;\beta_2\beta_1}(\omega), \label{eq:Magnetization}\\
            \langle S^{\alpha_1}_i S^{\alpha_2}_j\rangle &= \frac{1}{\beta}\sum_\omega\chi^{\alpha_2\alpha_1}_{ij}(\omega), \label{eq:EqTimeSus}
\end{align}
\begin{widetext}
\begin{align}
    \chi^{\alpha_1\alpha_2}_{ij}(\omega) = &
    \beta\delta_{0,\omega}M^{\alpha_1}_i M^{\alpha_2}_j
    +
    \delta_{ij}\sum_{\substack{\omega_1 \\\alpha\beta\gamma\delta}}
   \frac{\epsilon_{\alpha_{2}\beta_{1}\beta_{2}}\epsilon_{\alpha_{1}\beta_{3}\beta_{4}}}{4}
    \left[G^T_{i;\beta_{4}\beta_{1}}(\omega_1)G^T_{i;\beta_{2}\beta_{3}}(\omega_1+\omega)
    -
    G^T_{i;\beta_{3}\beta_{1}}(\omega_1)G^T_{i;\beta_{2}\beta_{4}}(\omega_1+\omega)\right] \label{eq:Susceptibility}\\
    +&
    \sum_{\substack{\omega_1 \omega_2 \\\beta_1\ldots\beta_4 \\\gamma_1 \ldots \gamma_4}}
    \frac{\epsilon_{\alpha_{2}\beta_{1}\beta_{2}}\epsilon_{\alpha_{1}\beta_{3}\beta_{4}}}{4} 
   G_{\beta_{4}\gamma_{4}}^{T}(\omega_{1}-\nu)G_{\gamma_{3}\beta_{3}}^{T}(\omega_{1})G_{\beta_{2}\gamma_{2}}^{T}(\omega_{2}+\nu)G_{\gamma_{1}\beta_{1}}^{T}(\omega_{2})\Gamma^T_{ij;\gamma_{4}\gamma_{3}\gamma_{2}\gamma_{1}}(-\nu,-\omega_{1}-\omega_{2},\omega_{2}+\nu-\omega_{1})
\end{align}
\end{widetext}

To verify the correctness of our implementation, in  App.~\ref{app:Dimer}, we consider a simple, exactly solvable model of two interacting spins. Note that this model poses the same methodological challenge to our method as infinite systems and thus provides an excellent benchmark. Overall, we observe similar or better results as compared to the $\Lambda$-flow method. Note that as detailed in the App.~\ref{app:Consistency}, other checks via exact relations between vertices are also possible, but less reliable as they check only for conservations of specific constants of motions which may be unrelated to quantities of interest.
\section{\texorpdfstring{$J_1$-$J_2$}{} Cubic lattice Heisenberg Model}
\subsection{Magnetic phase diagram}
\label{sec:J1J2}
\begin{figure}
    \centering
    \includegraphics[width = \linewidth]{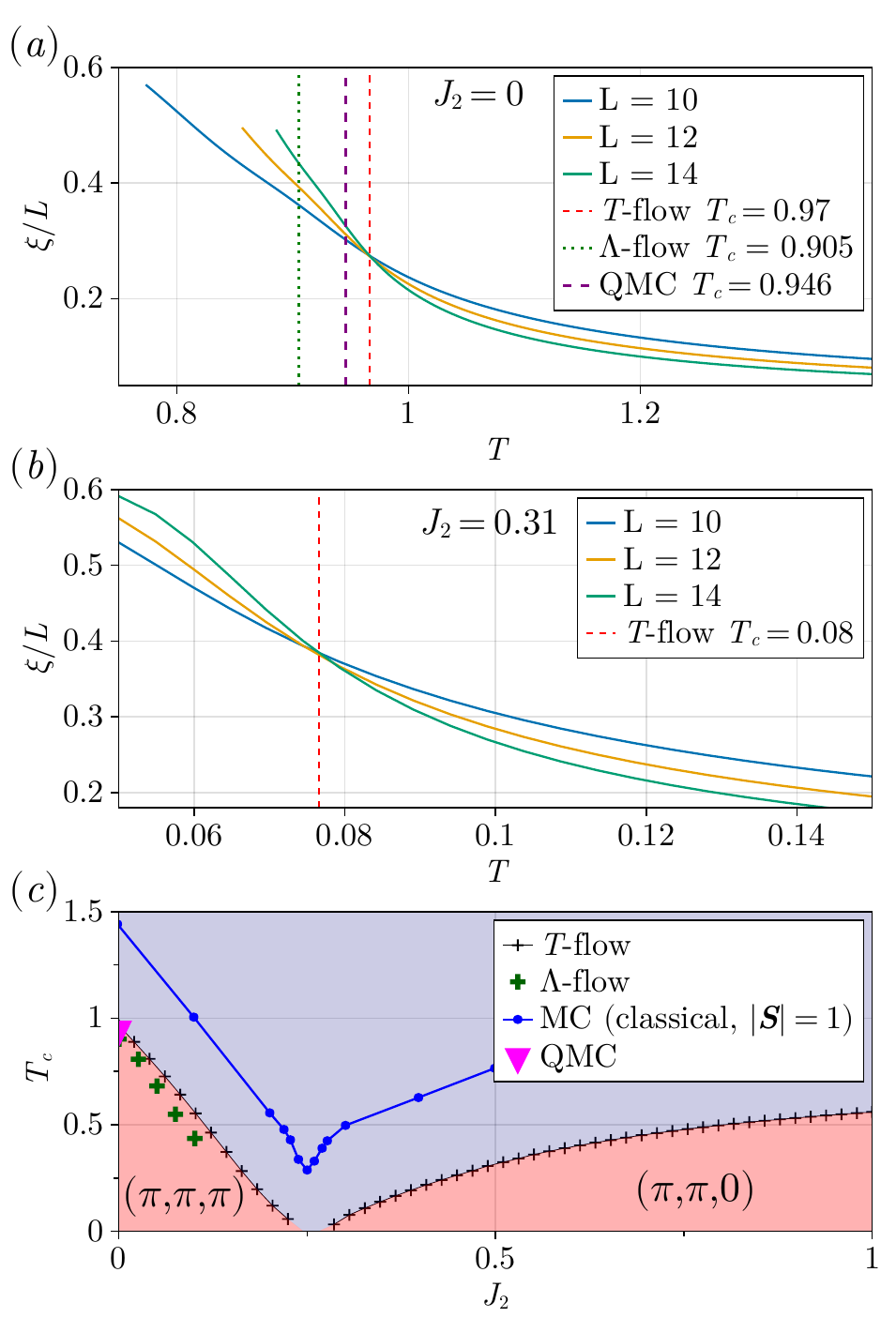}
    \caption{PMFRG results for the $J_1$-$J_2$ Heisenberg model on the cubic lattice. (a) Finite-size scaling of the correlation length using $T$-flow PMFRG for the simple cubic Heisenberg antiferromagnet at $J_2=0$ in comparison the standard $\Lambda$-flow PMFRG \cite{niggemann_quantitative_2022} and quantum Monte Carlo (QMC) \cite{Sandvik1998}.
    (b) Finite-size scaling of the correlation length using $T$-flow PMFRG for the simple cubic Heisenberg antiferromagnet at $J_2=0.31$.
    (c) Phase diagram: The transition temperature for the classical model with unit spin length is reproduced from Ref.~\cite{pinettes_phase_1998} (blue). The critical temperatures obtained from $T$-flow (black crosses) predicts slightly larger transition temperatures than the standard $\Lambda$-flow PMFRG (green) in the N\'eel ordered regime. At $J_2\gtrsim0.25$ the $T$-flow scheme detects critical scaling towards a stripe-ordered phase in qualitative agreement to the classical model. This critical temperature is not detected in the $\Lambda$-flow scheme.
    }
    \label{fig:CubicOverview}
\end{figure}

As a first non-trivial model to test our $T$-flow PMFRG approach, we revisit the antiferromagnetic $J_1$-$J_2$ Heisenberg model on the cubic lattice which was previously treated in the $\Lambda$-flow formalism \cite{niggemann_quantitative_2022}. Here, $J_{1}$ and $J_{2}$ are the antiferromagnetic exchange interactions on nearest- and next-nearest neighbor bonds, respectively,
\begin{align}
    H =  J_1 \sum_{\langle i,j\rangle ,\alpha} S^\alpha_i S^\alpha_j + J_2 \sum_{\langle\langle i,j\rangle\rangle,\alpha}  S^\alpha_i S^\alpha_j.
\end{align}
We set $J_1=1$ and first consider the case $J_2=0$. In this simple case, the model is unfrustrated and we expect a transition to antiferromagnetic (AFM) N\'eel order with wave vector ${\bf k} =(\pi,\pi,\pi)$ at some finite temperature that can be compared to quantum Monte Carlo. Fig.~\ref{fig:CubicOverview}(a) displays a critical scaling of the correlation length as indicated by the line crossings of $\xi/L$, where $L$ is the spatial cutoff distance beyond which vertices are approximated as zero. We detect the critical temperature $T_c \approx 0.97$, for details see App.~\ref{app:Scaling} and Refs.~\cite{schneider_taming_2022,muller2023pseudo,Sandvik1998}. Our result is in good agreement to quantum Monte Carlo ($T_c = 0.946$). Incidentally, and perhaps accidentally, we find this result to be marginally better compared to the conventional $\Lambda$-flow PMFRG which slightly underestimated the critical temperature as $T_c = 0.905$ \cite{niggemann_quantitative_2022}.

In the frustrated regime at finite $J_2>0$, 
Monte Carlo for classical spins ($|{\bf S}|=1$) \cite{pinettes_phase_1998} predicts order at finite temperatures throughout the phase diagram, with N\'eel order for $J_2<0.25$, which is replaced by antiferromagnetic stripe order with wave vector ${\bf k}=(\pi,\pi,0)$ (and symmetry related wave vectors) for $J_2>0.25$ as seen in Fig.~\ref{fig:CubicOverview}(c). 

In the quantum spin-1/2 case, the possible presence of a  small nonmagnetic region around $J_2=0.25$ is still debated. 
At $T=0$, linear spin wave theory \cite{majumdar_zero_2010} and the coupled cluster method \cite{farnell_ground-state_2016} predict antiferromagnetic order from $J_2=0$ that transitions into a small paramagnetic phase at $J_2\approx0.25$ before undergoing a second phase transition into the antiferromagnetic stripe phase for $J_1/J_2>0.25$. On the other hand, nonlinear corrections \cite{majumdar_zero_2010} to spin wave theory as well as a variational cluster approach \cite{laubach_quantum_2016}, predict no paramagnetic phase between the two ordered phases. 
Using our $T$-flow PMFRG as outlined above, we determine the finite temperature phase diagram, detecting critical temperatures down to a minimum simulation temperature of $T \sim 0.05$, with the case $J_2=0.31$ shown in Fig.~\ref{fig:CubicOverview}(b). 
The full phase diagram obtained this way is shown in Fig.~\ref{fig:CubicOverview}(c).
In agreement with other methods, we find a phase transition to antiferromagnetic N\'eel order for $J_2\lesssim 0.25$ and to antiferromagnetic stripe order for $J_2\gtrsim0.25$. Due to the observed critical scaling in system size, all phase transitions are of second order. In between, we observe a small regime without any sign of magnetic order.
Although intrinsic consistency checks seem to indicate less accurate results at lower temperatures (see App.~\ref{app:Consistency}), our findings support claims that there might be a small region with a paramagnetic phase in between the antiferromagnetic N\'eel and stripe ordered phases.  

\subsection{Discussion of the stripe phase transition}\label{sec:discussion_transition}
As discussed in the previous subsection, we detect a second order phase transition towards stripe order in the regime $J_2\gtrsim 0.25$. Although  expected from other methods, this result initially appears incompatible with previous findings in the $\Lambda$-flow scheme, which, despite observing large dominant stripe correlations could not detect a critical scaling \cite{niggemann_quantitative_2022}. 
We now show that this apparent discrepancy has a simple explanation by further including this artificial infrared cutoff $\Lambda$ into our temperature flow scheme and interpreting it as an auxiliary parameter.

To compare differences between the two flow schemes we dress the temperature flow propagator with the usual cutoff of the $\Lambda$-flow \cite{niggemann_frustrated_2021}, $\Theta^\Lambda(\omega) = \frac{T^2 \omega^2}{T^2\omega^2 + \Lambda^2}$ so that

\begin{equation}
     G^{-1,T}_{i;\alpha_1\alpha_2}(\omega_1) = \frac{1}{\Theta^\Lambda(\omega_1)}G^{-1,T}_{0;i;\alpha_1\alpha_2}(\omega_1) - \Sigma^T_{i;\alpha_1\alpha_2}(\omega_1). \label{eq:HybridG}
\end{equation}

\begin{figure}
    \centering
    \includegraphics[width = \linewidth]{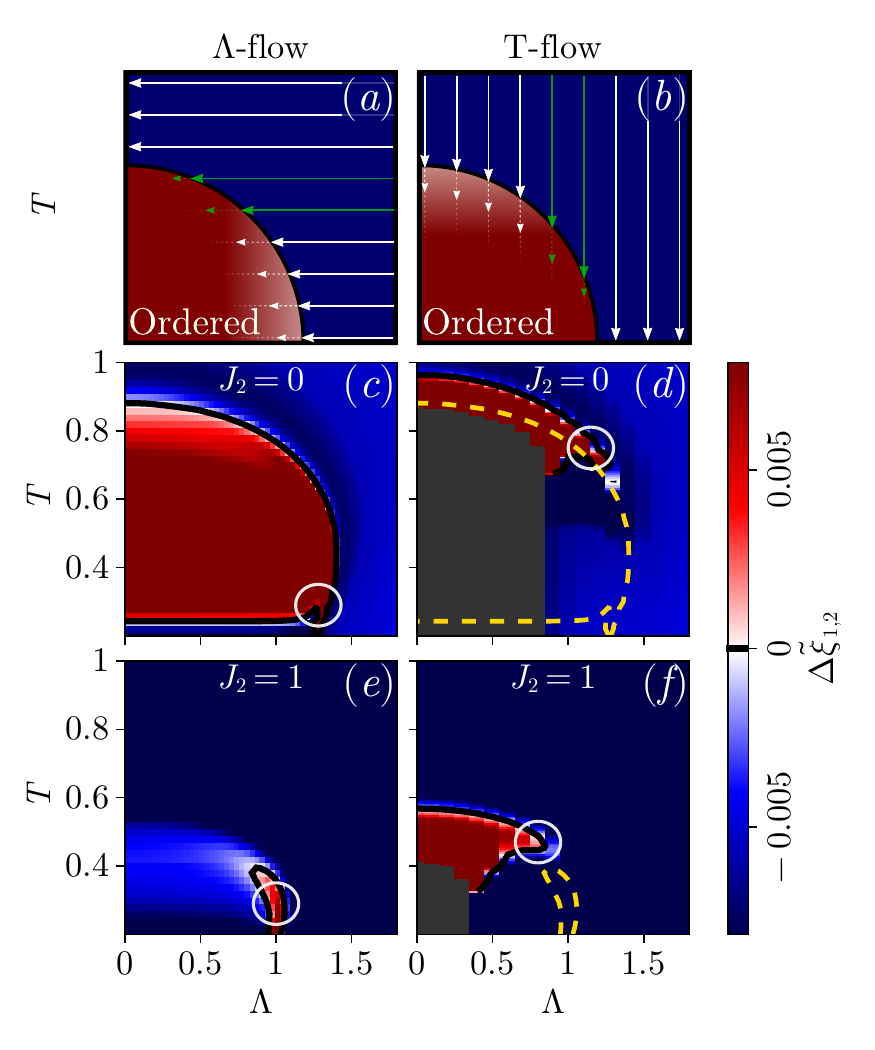}
    \caption{Magnetic phase 
    diagram for the $\Lambda$-flow scheme (left) and the $T$-flow scheme (right) both as functions of physical temperature $T$ and the artificial infrared cutoff $\Lambda$.
    (a,b): Schematic picture of the phase diagram and the direction of the flow (arrows) for both schemes. Each arrow represents an independent FRG run. Green arrows indicate problematic flow paths along the circumference of the ordered dome close to the phase boundary.
    (c-f): The ordered (paramagnetic) phase is determined by a positive (negative) difference of the rescaled correlation length in Eq.~\eqref{eq:rescaledXi} between two runs for $L_1=14$ and $L_2=12$. Additionally, the phase boundaries given by the contour $\Delta \tilde{\xi} = 0$ are indicated by black lines. In (d) and (f), the phase boundary from the $\Lambda$-flow scheme is displayed in yellow. White circles highlight exemplary positions of breakdowns of the PMFRG flow.
    For better visibility, the color range is limited to a small region around $\Delta \tilde{\xi} = 0$.
    } 
    \label{fig:flowScalingComparison}
\end{figure}

By construction, in the limit $\Lambda =0$, \cref{eq:HybridG} reduces to the propagator introdused in the previous section.
This propagator is now equal to the $\Lambda$-flow propagator in the entire $T,\Lambda$ parameter space (aside from the trivial pre\-factors of $T^{1/2}$ due to the rescaling of Majorana fields). Hence, physical observables at large $T$ or $\Lambda$ will be equal in both approaches. If both $T \lesssim 1$ and $\Lambda \lesssim 1$, however, the approximation of neglecting higher order vertices becomes uncontrolled, generally allowing for different results between the two methods.
\Cref{fig:flowScalingComparison} shows a comparison of the $T$-flow scheme (dressed with a $\Lambda$ cutoff) and the $\Lambda$-flow scheme as a function of $T$ and $\Lambda$ both at $J_2 = 0$ (N\'eel order) and $J_2 = 1$ (stripe order).  As both $T$ and $\Lambda$ suppress spin correlations, magnetic order can only be stabilized in a finite region around $T=\Lambda = 0$ as indicated schematically in Fig.~\ref{fig:flowScalingComparison}(a,b).
As displayed further, the conventional $\Lambda$-flow scheme approaches the ordered phase along lines of constant $T$ while the $T$-flow approaches it along constant $\Lambda$.

The remaining panels (c-f) display the difference of the rescaled correlation length for the dominant susceptibility
\begin{equation}
    \Delta \tilde \xi_{1,2} = \frac{\xi(L_1)}{L_1} - \frac{\xi(L_2)}{L_2} \label{eq:rescaledXi}
\end{equation} 
for two different spatial cutoff distances $L_1 > L_2$. At the phase transition, we have $\xi(L) \propto L$ and thus $\Delta \tilde \xi = 0$. Consequently, for large enough $L_{1,2}$, we can identify the region with $\Delta \tilde \xi >0$ ($\Delta \tilde \xi <0$) as the ordered (disordered) phase.

For $J_2=0$ [see Fig.~\ref{fig:flowScalingComparison}(c) and (d)] both $\Lambda$ and $T$-flows find magnetic order at $\Lambda=0$ for $T\approx0.9$. Although RG flows can become unphysical below the critical scale of a phase transition, in the $\Lambda$-flow the susceptibility and correlation lengths converge to a large but finite plateau value. For small temperatures of $T < 0.3$, on the other hand, we observe a very different behavior of the correlation length which displays a peak as a function of $\Lambda$ at a finite $\Lambda \sim 1.25$ indicated by the white circle in Fig~\ref{fig:flowScalingComparison}(c).  This sharp feature, also referred to as a \emph{flow breakdown}, originates from a peak of the maximum susceptibility (see App.~\ref{app:Scaling})
in the renormalization flow. In zero-temperature approaches it is an established signature of a phase transition \cite{muller2023pseudo,reuther_j_2010,iqbal2019quantum,iqbal2016functional,muller2023pseudo}, whose detection, however, can be ambiguous in practice.

Below the critical temperature, the $T$-flow correlations grow rapidly. Numerically, this requires increasingly smaller steps when solving the flow equations, which we eventually terminate as seen for $J_2=0$ in Fig.~\ref{fig:flowScalingComparison}(d). Strikingly, at a finite value of $\Lambda$ in Fig.~\ref{fig:flowScalingComparison}(d), the scaling collapse is no longer obtained, leaving the right boundary of the magnetic phase seemingly absent, with similar flow breakdown features as found in the $\Lambda$-flow scheme, also indicated via a white circle.

We now move on discussing the $T$-$\Lambda$ phase diagrams at $J_2=1$ for both, the $\Lambda$-flow and $T$-flow schemes in Fig.~\ref{fig:flowScalingComparison}(e) and (f), respectively. The $T$-flow result in Fig.~\ref{fig:flowScalingComparison}(f) resembles the observation in Fig.~\ref{fig:flowScalingComparison}(d) in that a critical scaling is only found at small $\Lambda$ but disappears as $\Lambda$ increases. The $\Lambda$-flow behavior at $J_2=1$ in Fig.~\ref{fig:flowScalingComparison}(e) also resembles Fig.~\ref{fig:flowScalingComparison}(d) and (f) {\it but with the roles of $T$ and $\Lambda$ reversed:} While Fig.~\ref{fig:flowScalingComparison}(e) only displays a phase transition at finite $\Lambda \sim 1$ and small $T\lesssim 0.3$, critical scaling is never found in the physically relevant limit $\Lambda = 0$. This makes it impossible to extract a critical temperature in the $J_2>0.25$ parameter regime within the $\Lambda$-flow scheme.

We interpret these results as follows: Clearly, both, $\Lambda$-flow and $T$-flow PMFRG methods are sensitive to ordering tendencies. However, each approach is better suited to detect phase boundaries that do not require a long flow through a critical region close to a magnetic phase. For example, such situations occur when a magnetically ordered phase is only grazed during the renormalization group flow in either $\Lambda$ or $T$,  shown by green arrows in panels (a) and (b) of \cref{fig:flowScalingComparison}.
In these critical regions, vertices grow large and the approximation of neglecting higher-order vertices is no longer accurate. Concretely, this means that the $\Lambda$-flow scheme is more sensitive to phase boundaries found at finite $\Lambda$, while the $T$-flow is better at detecting the opposite boundary at finite $T$ and small values of $\Lambda$. 
Indeed, one can approximate the shape of the full magnetic phase in the $T$-$\Lambda$ space by the complement of both methods. This is visualized by the yellow dashed line in Fig.~\ref{fig:flowScalingComparison}(d) and (f).

We see that for $J_1 = 1$ the challenge to resolve magnetic order in the $\Lambda$-flow scheme is especially pronounced as the top phase boundary in $T$-$\Lambda$ space is particularly flat [see Fig.~\ref{fig:flowScalingComparison}(f)], and the ordering temperature much smaller.  To extract physical quantities, the FRG in the $\Lambda$-flow scheme needs to be solved all the way down to $\Lambda = 0$, possibly flowing through an ordering transition, where the truncation of flow equations is known to break down. The temperature flow on the other hand needs to be followed only slightly beyond the boundary of the phase transition. We can therefore conclude that the $T$-flow scheme is the favorable method as it approaches the phase boundary from the physically relevant direction.

\section{Square lattice dipolar XXZ model}
\label{sec:Dipolar}
In this section, we present another application of the $T$-flow PMFRG, demonstrating that this approach is capable to treat a complex long-range interacting two-dimensional spin model of current research interest and to accurately capture its finite temperature ordering transitions. 

\begin{figure}
    \centering
    \includegraphics[scale=1.0]{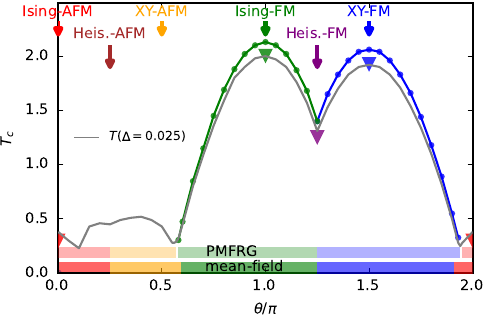}
    \caption{Finite temperature phase diagram for the square lattice dipolar XXZ model of Eq.~\eqref{eq:dipolar}. Dots denote the critical temperature found from finite-size scaling of the $T$-flow PMFRG susceptibility. Lines are guides to the eye. The gray line denotes the lowest accessible temperature below which the internal consistency check for the PMFRG is violated by $2.5\%$, see App.~\ref{app:Consistency}. The color of the upper horizontal bar denotes the dominant susceptibility (Ising-AFM in red, XY-AFM in orange) at this lowest temperature. The lower bar shows the $T=0$ mean-field phase boundary determined in Ref.~\cite{peter_anomalous_PhysRevLett.109.025303} and the arrows with labels on top denote special cases of the model in Eq.~\eqref{eq:dipolar}. The triangles indicate critical temperatures from classical and quantum Monte Carlo simulations of Refs.~\cite{zhaoFinitetemperatureCritical2023,sbierski2023magnetism,rastelliPhaseTransitions2006} and from App.~\ref{app:classMC}.
    }
    \label{fig:dipolarTc}
\end{figure}

If an ordered phase in two dimensions breaks a continuous (spin-rotation) symmetry, a finite $T_c$ is only possible for sufficiently long-ranged interactions \cite{Mermin_Absence_PhysRevLett.17.1133, defenu_longRange_RevModPhys}. Dipolar interactions, decaying with distance as $1/r^{3}$ are such an example of experimental relevance as they can be realized in systems of cold atoms and molecules \cite{chomazDipolarPhysics2022,christakisProbingSiteresolved2023} or Rydberg atom arrays \cite{chenContinuousSymmetryBreaking2023}. Here, following the early work of Peter et al.~\cite{peter_anomalous_PhysRevLett.109.025303}, we focus on the spin-1/2 square lattice XXZ model with isotropic dipolar interactions and $U(1)$ spin rotation symmetry,
\begin{equation}
    H=\sum_{(i,j)}J_{(i,j)}^{x}\left(S_{i}^{x}S_{j}^{x}+S_{i}^{y}S_{j}^{y}\right)+J_{(i,j)}^{z}S_{i}^{z}S_{j}^{z} \label{eq:dipolar}
\end{equation}
where the sum is over all bonds of the square lattice and $J_{(i,j)}^{x}=\sin(\theta)/r_{ij}^{3}$, $J_{(i,j)}^{z}=\cos(\theta)/r_{ij}^{3}$ with $r_{ij}=|\mathbf{r}_i-\mathbf{r}_j|$ and $\mathbf{r}_i$ is the position of lattice site $i$. The angle $\theta \in [0,2\pi)$ controls the ratio between Ising and in-plane interactions and interpolates between the special cases of AFM-Ising ($\theta=0$), XY-AFM ($\theta=\pi/2$), Ising-FM ($\theta=\pi$) and XY-FM model ($\theta=3\pi/2$), see Fig.~\ref{fig:dipolarTc}. 

For $J_x\leq0$ ($\theta\in [\pi,2\pi]$) the model is free of the sign problem and thus amenable to quantum Monte Carlo simulation. Results exist for critical temperatures of the Heisenberg-FM and the XY-FM case, $\theta/\pi=1.25$ and $1.5$ \cite{zhaoFinitetemperatureCritical2023,sbierski2023magnetism}. For the Ising cases at $\theta=0,\pi$, classical Monte Carlo simulations are applicable. For the Ising-AFM case, $T_c=0.296$ was found~\cite{rastelliPhaseTransitions2006}. For the Ising-FM case, where no classical Monte Carlo results were available in the literature, we obtained $T_c=2.00(1)$. Details are reported in \cref{app:classMC}. All classical and quantum Monte Carlo results are shown in Fig.~\ref{fig:dipolarTc} as triangles.  

In order to complete the phase diagram and find ordering temperatures for all $\theta$ (if non-zero and large enough to detect with PMFRG), we apply the PMFRG $T$-flow method and report the results of a finite-size scaling procedure in Fig.~\ref{fig:dipolarTc} (blue and green dots). Our PMFRG results are usually a few percent above Monte Carlo estimates where the latter are available. The $T_c(\theta)$ corresponding to either Ising- or XY-type FM order forms two domes merging at the Heisenberg FM point ($\theta=1.25 \pi$) at which the $T_c$ is sharply suppressed. Extrapolating the $T_c(\theta)$ to vanishing temperature we find that the ordered regions are slightly extended beyond the mean-field ground-state phase-boundaries of Ref.~\cite{peter_anomalous_PhysRevLett.109.025303}. The latter are shown by a colored bar in the bottom of Fig.~\ref{fig:dipolarTc}.

In the region with dominant AFM interaction, the PMFRG did not find an ordering transition for ${\theta/\pi=1.94...0.57}$ down to the lowest accessible temperatures marked by the gray line. This lower bound corresponds to the temperature below which the internal consistency check is violated by $2.5\%$, see App.~\ref{app:Consistency}. To indicate the type of (short-range) spin correlations in the $\theta$-region where no order is detected, the color of the upper vertical bar specifies the dominant susceptibility (red for Ising-AFM, orange for XY-AFM) at this lowest temperature. In the region around the Ising-AFM point where the order breaks a discrete global spin-flip symmetry, a finite $T_c$, albeit small due to frustration, is expected. Indeed, as mentioned above, classical Monte Carlo~\cite{rastelliPhaseTransitions2006} finds $T_c\simeq 0.296$ which is just below the temperature accessible to PMFRG. In the region around the XY-AFM point, a Kosterlitz-Thouless transition is anticipated at finite $T$ with true long-range order only appearing at $T=0$, see Ref.~\cite{peter_anomalous_PhysRevLett.109.025303,chenContinuousSymmetryBreaking2023} for further discussion. The PMFRG currently is not able to detect the Kosterlitz-Thouless transition, a challenge remaining for future work. 

For all $\theta$, we find the dominant susceptibilites to be either of XY or Ising type and located at $\mathbf{k}=0$ or $(\pi,\pi)$ as indicated in Fig.~\ref{fig:dipolarTc} by the upper horizontal line.  In particular, this means that we do not find any signs of exotic magnetic (e.g.~incommensurate) phases. Moreover our results do not rule out paramagnetic behavior down to $T=0$ which is conceivable close to the mean-field phase boundaries at $\theta/\pi\simeq 0.6$ and $1.9$ where the mean-field energies are discontinuous~\cite{peter_anomalous_PhysRevLett.109.025303}.
In summary, our results inform future experiments which could map out the two-dome structure of $T_c(\theta)$ and further explore the nature of the non-magnetic low temperature states with dominant XY-AFM interactions. The PMFRG is also capable of treating the case of tilted dipoles which results in anisotropic spin interactions.

\section{Conclusion}
\label{sec:Conclusion}

Motivated by the efficiency of using a physical flow parameter, in this work we have implemented the temperature flow into the PMFRG framework. Benchmarking our method on the nearest neighbor simple cubic Heisenberg model we find this new method to have similar or better accuracy as compared to the standard $\Lambda$-flow formalism. For the Heisenberg $J_1$-$J_2$ model on the simple cubic lattice, we resolve a previous problem about the inability to detect critical scaling at $J_2>0.25$ in a $\Lambda$-flow study~\cite{niggemann_quantitative_2022}. Our explanation indicates that unphysical renormalization cutoff parameters $\Lambda$ can suffer from the onset of long-range correlation effects at finite cutoff values, thereby rendering the result in the physical limit at $\Lambda=0$ inaccurate. The use of the temperature flow also allows us to detect magnetic order at very low critical temperatures $T \sim 0.05$ that were previously out of reach.
We demonstrate the substantial improvement in efficiency of the $T$-flow PMFRG by mapping out the phase diagrams of the Heisenberg $J_1$-$J_2$ model on the simple cubic lattice and the dipolar-XXZ model on the square lattice. Both models have points in parameter space where they are amenable to (quantum) Monte Carlo, allowing us to verify the quantitative accuracy of our approach up to a few percent, which gives us confidence in the quantitative accuracy of our critical temperatures even in cases where no Monte Carlo benchmark is available. Combining the efficiency of the zero temperature $\Lambda$-flow approach with the methodological advantages of the finite temperature formalism, we strongly encourage the use of temperature as the preferred flow parameter.

\section*{Acknowledgements}
We thank Lode Pollet for stimulating discussion. We further acknowledge fruitful discussions with the other participants from the spin-FRG mini-workshop 2023 in Berlin during the early stages of this project.
N.~N.~and J.~R.~acknowledge support from the Deutsche
Forschungsgemeinschaft (DFG, German Research Foundation), within Project-ID 277101999 CRC 183 (Project
A04).
B.~Sch.~and B.~Sb.~are supported by a MCQST-START fellowship and by
the Munich Quantum Valley, which is supported by the
Bavarian state government with funds from the Hightech
Agenda Bayern Plus. B.~Sb.~acknowledges support from DFG through the Research Unit FOR 5413/1,
Grant No.~465199066. 
We acknowledge the use of the JUWELS cluster at the Forschungszentrum J\"ulich, the Noctua2 cluster at the Paderborn Center for Parallel Computing (PC$^2$), and the HPC Service of ZEDAT at the Freie Universität Berlin.

\appendix

\section{Heisenberg dimer as benchmark model for $T$-flow PMFRG}
\label{app:Dimer}
\begin{figure}
    \centering
    \includegraphics[width = \linewidth]{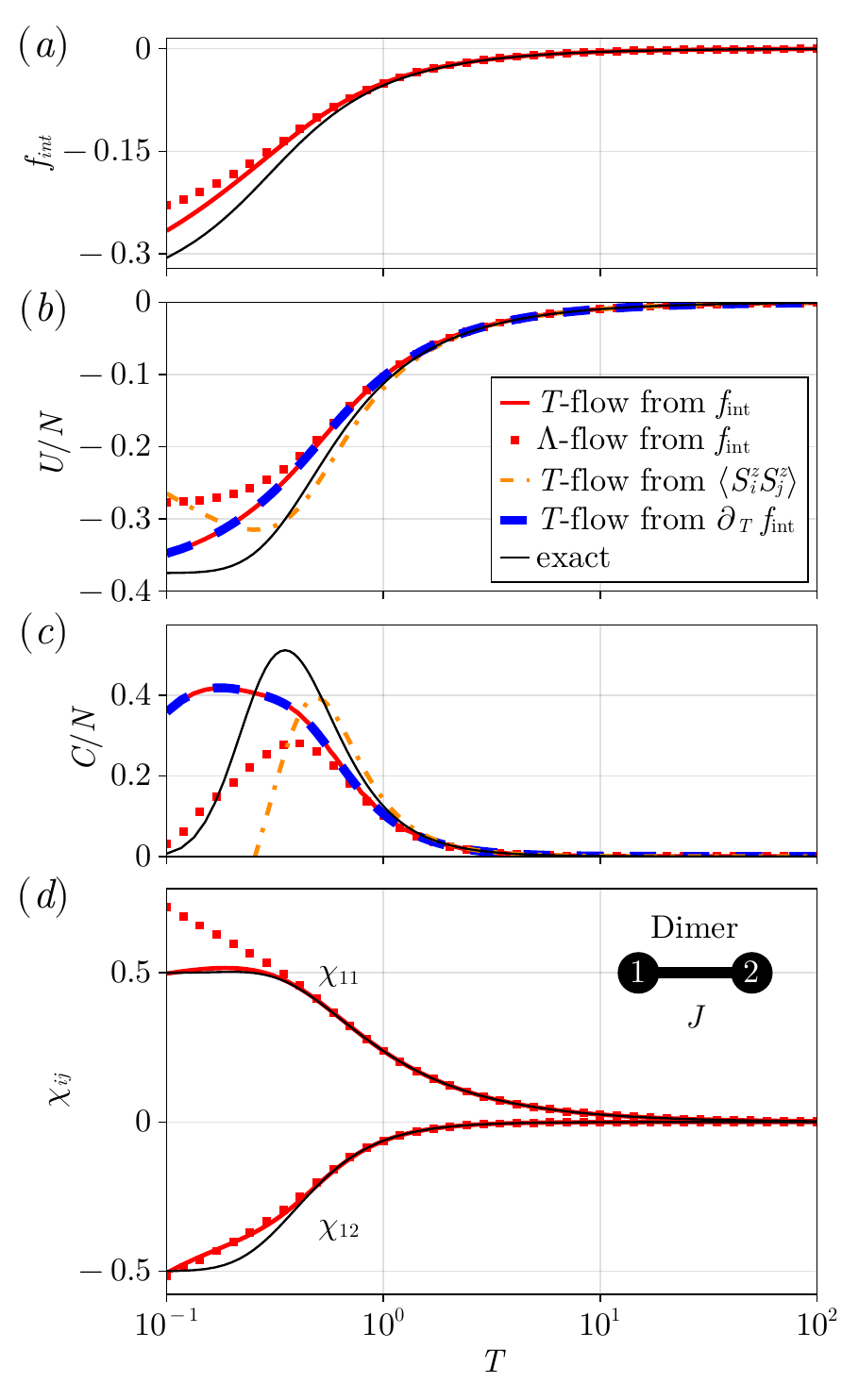}
    \caption{Thermodynamic quantities for the temperature flow PMFRG on the Heisenberg dimer in comparison to the exact result (black) and the standard $\Lambda$-flow PMFRG (squares). (a): interaction correction to the free energy $f_\textrm{int}$ from \cref{eq:FlowEqG0}. For the energy per site $U/N$ (b) and the specific heat $C/N$ (c), the darkred dash-dotted line represents the value obtained via Eq.~\eqref{eq:e_chi}. The solid red line represents the value obtained via Eq.~\eqref{eq:HeatCapMain}. The same quantity can be obtained directly via the flow equation Eq.~\eqref{eq:FlowEqG0} shown in blue dashed lines without the need to perform numerical derivatives. (d) shows the two inequivalent static spin-spin correlators $\chi_{11}$ and $\chi_{12}$ obtained via \cref{eq:Susceptibility}.}
    \label{fig:ThermoDimer}
\end{figure}

To benchmark the temperature flow PMFRG, we investigate the Heisenberg dimer $H = \sum_{\alpha} S^\alpha_1 S^\alpha_2$. The same system has been studied previously for similar purpose \cite{niggemann_frustrated_2021,schneider_taming_2022}.
Here, we consider the static spin-spin correlators $\chi_{11}(\omega=0)$ and $\chi_{12}(\omega=0)$ as well as the interaction correction to the free energy $f_\textrm{int}$, the energy per site $U$ and the heat capacity $C$ obtained by Eqs.~\eqref{eq:FreeEnergy}-\eqref{eq:HeatCapMain}.
Alternatively, the internal energy can also be obtained via $U=\langle H\rangle$ which for the general Hamiltonian in Eq.~(\ref{eq:ham_general}) reads as
\begin{equation}
    U = \sum_{i,\alpha} h^\alpha_i M^\alpha_i + \frac{1}{2} \sum_{\substack{i,j\\ \alpha \beta}} J^{\alpha,\beta}_{ij} \langle S^\alpha_iS^\beta_j\rangle, \label{eq:e_chi}
\end{equation}
where $M_i^\alpha=0$ in the present case, since no magnetic field $h_i^\alpha$ is considered. These quantities are compared against the exact solution in Fig.~\ref{fig:ThermoDimer} shown as black lines: The interaction correction to the free energy $f_\textrm{int}$, shown in red in panel (a), is obtained from the zero-point vertex in \cref{eq:FlowEqG0}. We observe the temperature flow (solid line) to be closer to the exact result than the $\Lambda$-flow result (square markers). From $f_\textrm{int}$, the energy per site $U/N$ may be obtained using \cref{eq:MeanEnergy} via a numerical derivative with respect to $T$. Again, we observe the $T$-flow curve to be closer to the exact result than in $\Lambda$-flow in panel (b). In the $T$-flow scheme, we may avoid inaccuracies from numerical derivatives by inserting the right hand side of the flow equation in Eq.~(\ref{eq:FlowEqG0}) for $\frac{df_\textrm{int}}{dT}$ in \cref{eq:MeanEnergy}. The result is shown by the blue dashed line. As the numerical accuracy of the solution is rather high with a tolerance of $\sim 10^{-7}$, the result is identical to the one obtained via numerical derivatives.
Further shown in orange is the $T$-flow result obtained from spin-spin correlations as defined in \cref{eq:e_chi}. For intermediate to large temperatures, this quantity is the most accurate but becomes unphysical around $T\sim 0.25$, showing an increase as the temperature decreases.
By taking a numerical derivative, we may also obtain an estimate for the heat capacity $C$ from all these results, shown in panel (c). While the $T$-flow peak height of the heat capacity is closer to the exact result than the $\Lambda$-flow result, its peak location is shifted. We conclude that the energy per site and the heat capacity are strongly affected by truncation errors, since already small errors introduced by neglecting the six-point vertex propagate through the four- and two-point vertex to the zero-point vertex and are then magnified even further upon taking derivatives.

On the other hand, the static spin-spin correlations $\chi_{11}(\omega=0)$ and $\chi_{12}(\omega=0)$ are significantly less affected by this problem as they are obtained directly from the four-point vertex via Eq.~\eqref{eq:Susceptibility}. At large temperatures $T\gg J$, where both PMFRG flow approaches are well controlled, they agree well with each other and the exact result. At low temperatures, deviations from the exact result become visible within both the $T$-flow and standard $\Lambda$-flow PMFRG. Somewhat surprisingly, we observe that the local spin correlator $\chi_{11}(\omega=0)$ appears much more accurate in the temperature flow formalism, while the non-local one deviates from the exact result in the same way as in the $\Lambda$-flow scheme.

\section{Internal consistency checks for PMFRG}
\label{app:Consistency}
The truncation of the flow equation hierarchy by neglecting the six point vertex is an inherently uncontrolled approximation at low temperatures, making estimates of the exact error bars impossible. Instead, we can rely upon the fulfillment of a Ward identity as a qualitative measure of the truncation error to indicate challenging parameter regimes:
All pseudo-Majorana Hamiltonians feature a set of local constants of motion, 
\begin{equation}
    \theta_j = -2 i \eta^x_j\eta^y_j \eta^z_j.
\end{equation}
This allows to derive an exact relation between fully local two- and four-point Majorana correlators \cite{niggemann_frustrated_2021,Shnirman2003,Schad2015}.
Hence we may express the static the local spin-spin correlator which is usually computed from the four-point Majorana vertex [see \cref{eq:Susceptibility}] alternatively through the two-point Green function, here shown for the static part at $\omega=0$,
\begin{align}
    \chi^{\alpha_1\alpha_2}_{jj}(\omega = 0) = \sum_{\omega'}\frac{i}{\omega' \sqrt{T}} G^T_{j;\alpha_1\alpha_2}(\omega'). \label{eq:chi_local_G}
\end{align}

This relation must be satisfied for any exact calculation. For the approximate PMFRG, we can use the degree of violation as an internal consistency check and define the quantity

\begin{align}
    \Delta =  \abs{\frac{\chi^{\alpha_1\alpha_2}_{jj}(0)_1-\chi^{\alpha_1\alpha_2}_{jj}(0)_2}{\chi^{\alpha_1\alpha_2}_{jj}(0)_1 + \chi^{\alpha_1\alpha_2}_{jj}(0)_2}},  \label{eq:Ward}
\end{align}
where the subscripts 1 and 2 refer to the two different methods of computing $\chi^{\alpha_1\alpha_2}_{jj}(\omega = 0)$, via \cref{eq:chi_local_G} and \cref{eq:Susceptibility}, respectively. 

\begin{figure}
    \centering
    \includegraphics[width = \linewidth]{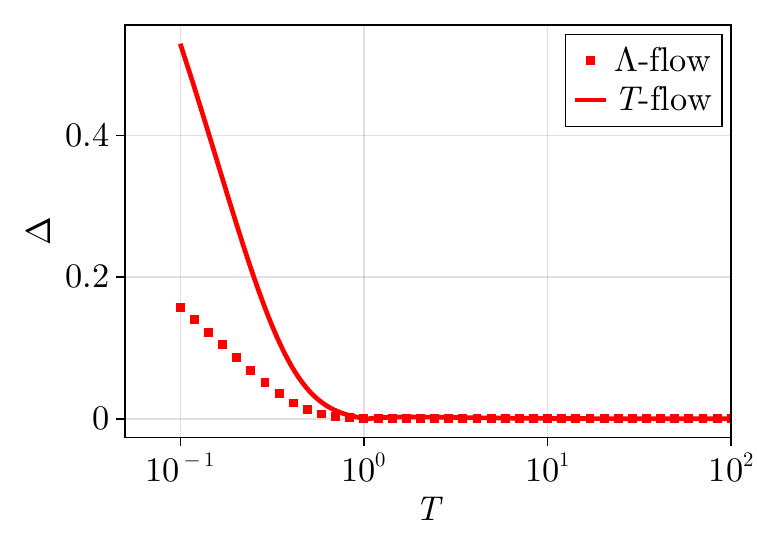}
    \caption{Violation of the consistency check [Eq.~(\ref{eq:Ward})] for the Heisenberg dimer in the $\Lambda$-flow and $T$-flow PMFRG schemes.}
    \label{fig:WardDimer}
\end{figure}

In Fig.~\ref{fig:WardDimer} we show the violation of the consistency condition $\Delta=0$ for the Heisenberg dimer from App.~\ref{app:Dimer} in the $\Lambda$ and $T$-flow schemes. We notice that $\Delta$ is larger in the temperature flow scheme, compared to the $\Lambda$-flow PMFRG, despite the overall better agreement of the temperature flow with the exact result.

\begin{figure}
    \centering
    \includegraphics[width = \linewidth]{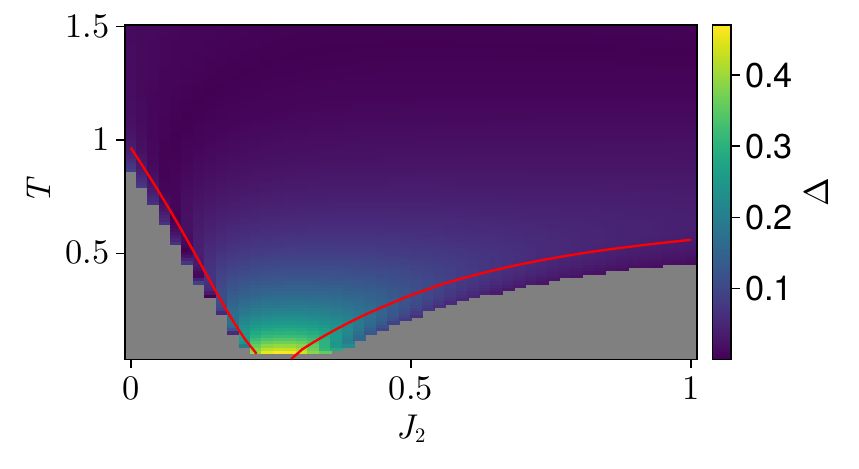}
    \caption{Consistency check violation $\Delta$, defined by \cref{eq:Ward}. The critical temperature is indicated by the red line.}
    \label{fig:Ward-identity_J2}
\end{figure}

\Cref{fig:Ward-identity_J2} shows the violation of the consistency check \cref{eq:Ward} for the $J_1$-$J_2$ Heisenberg model on the simple cubic lattice as a function of temperature and $J_2$ for the $T$-flow scheme with the critical temperatures in red corresponding to \cref{fig:CubicOverview}. At temperatures below $T \sim 0.6$, we find violations of about $5\%$, steadily growing towards lower temperatures up to $50\%$ at the very lowest temperatures $T=0.05$. 
Finally, for the XXZ dipolar model of Sec.~\ref{sec:Dipolar}, the temperature at which $\Delta=0.025$ is shown as a gray line in the phase diagram of Fig.~\ref{fig:dipolarTc}.

Although $\Delta$ can not replace a real error bar since it only contains information about the violation of the conservation law for the constant of motion $\theta_j$, a small value of $\Delta$ in the few-percent range is an indicator that the truncation of flow equations is still in the well-controlled limit.
However, it should be noted that even with a large $\Delta$ the method can produce qualitatively and in principle even quantitatively accurate data for quantities which are not directly linked to the conservation of $\theta_j$ which is violated. This is visible in the case of the dimer shown in \cref{fig:WardDimer} where $\Delta$ is larger in the temperature flow scheme as compared to the $\Lambda$-flow result even though the quantities of interest, primarily the susceptibility, lie closer to the exact result in $T$-flow.

\section{Detection of magnetic phase transitions}
\label{app:Scaling}
 
When studying spin systems at finite temperatures one is commonly interested in phase transitions or the lack thereof. Historically, magnetic phase transitions in the pseudoparticle-FRG context have been detected as instabilities in the flow equations, where a divergence is often detected as a sharp feature, i.e.~``kink'' in the corresponding susceptibility.
 This approach has the disadvantage that the exact point of the feature can heavily depend on numerical parameters such as the maximum correlation length, the frequency discretization or the accuracy of the ODE solver. Moreover, the distinction a weak ``kink'' from a disordered state is subject to interpretation and thus often of more qualitative nature. As outlined in previous works \cite{niggemann_quantitative_2022,schneider_taming_2022,sbierski2023magnetism}, finite-size scaling can instead be used as an unbiased and reliable method to extract quantitatively accurate critical temperatures from pseudo-particle-FRG calculations. We approximate the rescaled correlation length  by fitting a Lorentz curve with width $\frac{1}{\xi}$ to the largest peak  located at wavevector $\mathbf{Q}$ of the Fourier transformed susceptibility $\chi^{\alpha_1\alpha_2}(\mathbf{k})$\cite{sandvikComputationalStudiesQuantum2010}
\begin{align}
    \frac{\xi}{L} &= \frac{1}{2\pi}\max_{\boldsymbol{\delta}}\left(\sqrt{\frac{\chi_\textrm{max}(\mathbf{Q})}{\chi_\textrm{max}(\mathbf{Q}+\frac{2\pi}{L}\boldsymbol{\delta})}-1}\right)\\
    \chi_\textrm{max}(\mathbf{Q}) &= \max_{\alpha_1\alpha_2} \left(\chi^{\alpha_1\alpha_2}(\mathbf{Q})\right).
\end{align}
Here, $\boldsymbol{\delta}$ is a vector of unit length and $L$ is a measure of system size, and therefore, the maximum correlation length. In translationally invariant systems we need only consider sites $i$ in $\Sigma_i$ and $\Gamma_{ij}$ that lie in the first unit cell and set $\Gamma_{ij}=0$ if the sites $i$ and $j$ are separated by more than $L$ nearest neighbor bonds. 
We detect a phase transition by calculating $\frac{\xi}{L}$ for multiple $L$. In a paramagnetic regime $\frac{\xi}{L}$ decreases with $L$ while in a magnetic regime $\frac{\xi}{L}$ increases with $L$. The critical temperature is the temperature at which $\frac{\xi}{L}$ is independent of $L$.

\section{Numerical implementation}
Our numerical code is available on Github \url{https://github.com/NilsNiggemann/PMFRG.jl/tree/TemperatureFlow}, where implementations of the $T$-flow and $\Lambda$-flow can be found \cite{JuliaLambdaFlowPMFRG}.
\section{Classical Monte Carlo for the square lattice dipolar Ising model}
\label{app:classMC}
\begin{figure}
    \centering
    \includegraphics[width = \linewidth]{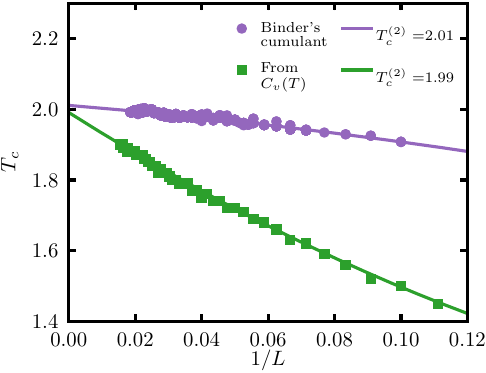}
    \caption{Scaling of the critical temperature extracted from the specific heat (green squares) and Binder's cumulant (purple circles). The lines correspond to quadratic fits and the results of $T_c$ for each method are shown in the legend.}
    \label{fig:cMC}
\end{figure}
We performed classical Monte Carlo calculations for the ferromagnetic dipolar Ising model [see Eq.~(\ref{eq:dipolar}) for $\theta = \pi$]. We use square systems with periodical boundary conditions containing $N=L\times L$ Ising spins, taking $L$ from 8 and up to 65. Each spin of the system interacts with all other spins through the exchange interaction $J_{(i,j)}=1/r_{ij}^{3}$, where $r_{ij}$ is the shortest distance between sites $i$ and $j$ on the torus.

To calculate the critical temperature, we use two independent methods. On one hand, we take 201 temperature steps to cool the system down from $T=3$ to 1 using $10^5$ Monte Carlo trials at each temperature. We measure the energy in the second half of each temperature step to obtain $U(T)/N$ and $c_v(T)$. Results are then averaged for 10 independent runs. On the other hand, we take 76 steps to cool down from $T=3$ to 1.5 using $2\times 10^5$ Monte Carlo trials at each temperature. We measure $m^2$ and $m^4$ in the same way as before and average over 10 independent runs to calculate Binder's cumulant~\cite{Binder81}, $B = (3-\langle m^4\rangle/\langle m^2\rangle^2)/2$. 

Fig.~\ref{fig:cMC} shows the temperature at which $c_v(T)$ has a maximum (green squares) and the crossing lines between Binder's cumulants for different lattice sizes (purple circles) as a function of $1/L$. Extrapolating to $L\rightarrow \infty$ via two independent quadratic fits yield $T_c = 1.99$ and $T_c=2.01$, respectively. Thus, we can assume that $T_c=2.00(1)$ is a good estimate of the critical temperature in the thermodynamic limit.

\bibliographystyle{apsrev4-1}
\bibliography{T-Flow}

\end{document}